\documentclass{ifacconf}

\usepackage{graphicx}
\usepackage{amsmath,amssymb,amsfonts}
\usepackage{bm}
\usepackage{natbib}        %
\usepackage{xcolor}
\usepackage{enumitem}
\usepackage{subcaption}
\usepackage{lipsum}

\newcommand{\N}[0]{\ensuremath{\mathbb{N}}}
\newcommand{\R}[0]{\ensuremath{\mathbb{R}}}
\newcommand{\Id}[0]{\ensuremath{\operatorname{Id}}}
\newcommand{\norm}[1]{\left\lVert#1\right\rVert}

\DeclareMathOperator*{\st}{subject~to}
\DeclareMathOperator*{\argmin}{arg\,min}

\makeatletter
\let\old@ssect\@ssect %
\makeatother

\usepackage{hyperref} %

\makeatletter
\def\@ssect#1#2#3#4#5#6{%
  \NR@gettitle{#6}%
  \old@ssect{#1}{#2}{#3}{#4}{#5}{#6}%
}

\makeatother
\begin{document}
\begin{frontmatter}

\title{Learning to accelerate Krasnosel'skii–Mann fixed-point iterations with guarantees\thanksref{footnoteinfo}}

\thanks[footnoteinfo]{This work was supported by Digital Futures and the Wallenberg AI, Autonomous Systems and Software Program (WASP) funded by the Knut and Alice Wallenberg Foundation.
}

\author[First]{Andrea Martin} 
\author[First]{Giuseppe Belgioioso}

\address[First]{Department of Decision and Control Systems, KTH Royal Institute of Technology, and Digital Futures, SE-100 44 Stockholm, Sweden (e-mail: \{andrmar, giubel\}@kth.se).}

\begin{abstract}                %
        We introduce a principled learning to optimize (L2O) framework for solving fixed-point problems involving general nonexpansive mappings. Our idea is to deliberately inject summable perturbations into a standard Krasnosel'skii–Mann iteration to improve its average-case performance over a specific distribution of problems while retaining its convergence guarantees. Under a metric sub-regularity assumption, we prove that the proposed parametrization includes only iterations that locally achieve linear convergence—up to a vanishing bias term—and that it encompasses all iterations that do so at a sufficiently fast rate. We then demonstrate how our framework can be used to augment several widely-used operator splitting methods to accelerate the solution of structured monotone inclusion problems, and validate our approach on a best approximation problem using an L2O-augmented Douglas–Rachford splitting algorithm. 
\end{abstract}
    

\end{frontmatter}

\section{Introduction}
Fixed-point iterations are a fundamental computational primitive in applied mathematics, engineering, and operations research. Among them, Krasnosel'skii–Mann (KM) iterations \citep{krasnosel1955two, mann1953mean} form a particularly important class, as they enable a unified convergence analysis of many algorithms—from decomposition methods in large-scale convex optimization \citep{ryu2022large} to equilibrium-seeking algorithms in multi-player games \citep{briceno2013monotone, belgioioso2022distributed}—through the lenses of monotone operator theory and nonexpansive mappings.

Despite their widespread use, however, standard KM iterations are known to be highly sensitive to ill conditioning and poor parameter selection, often resulting in slow convergence. This limitation has motivated the development of several acceleration schemes, which introduce momentum terms combining multiple past iterates \citep{walker2011anderson, bot2023fast, maulen2024inertial} and line-search strategies \citep{themelis2019supermann, giselsson2016line} to speed up convergence to high-accuracy solutions. Despite provably improving convergence in the worst case, these approaches may still be overly conservative, as the fixed-point problems encountered in practice rarely correspond to the adversarial instances for which the worst-case bounds are tight. In fact, many applications involve \emph{parametric families} of problems; in model predictive control \citep{rawlings2020model}, for instance, one computes the control action by repeatedly solving optimization problems that differ from one another only in the measured state. In these cases, the average solution time relative to a given distribution of problem parameters may offer a more representative measure of algorithmic performance than traditional worst-case convergence rates.

Motivated by this observation, the emerging paradigm of learning to optimize (L2O) \citep{chen2022learning, amos2023tutorial} adopts an alternative data-driven approach. Instead of hand-crafting acceleration schemes based on complex analyses of their worst-case rates, this paradigm uses machine learning (ML) to automate the design of solution methods tailored to the problem instances observed during training. By leveraging expressive neural network (NN) parametrizations to learn the structure underlying specific families of parametric problems, these methods can achieve significant speedups over traditional solvers, as recently shown in applications ranging from robotics \citep{sacks2022learning} to energy systems \citep{park2023compact}—see \cite{chen2022learning} for a comprehensive review.

However, while the flexibility of NN architectures enables learning powerful algorithms from data, directly replacing classical fixed-point iterations with learned updates often precludes the derivation of formal convergence guarantees \citep{chen2022learning}. To address this challenge, several L2O methods do not modify the analytic structure of classical fixed-point iterations, and only use ML to learn warm starts \citep{sambharya2024learning, celestini2024transformer} or hyperparameter values \citep{ichnowski2021accelerating}. To take full advantage of the expressiveness of NNs, another line of research seeks to learn parametrized update rules without imposing any analytic structure a priori. To ensure convergence in this setting, \cite{heaton2023safeguarded} introduced a safeguarding mechanism that reverts to a standard KM iteration whenever the learned updates become overly aggressive. In contrast, focusing on the case of linearly convergent operators, \cite{martin2025learning} bypasses the need for such fall-back mechanisms by parametrizing classes of iterations that guarantee linear convergence by design. This is achieved by characterizing perturbations to a base fixed-point iteration that preserve its global convergence guarantees—see also \cite{banert2024accelerated} and \cite{martin2024learning} for related results in the context of composite convex optimization and smooth nonconvex optimization, respectively. To the best of our knowledge, however, none of these approaches enables the application of L2O methods to fixed-point problems involving general nonexpansive mappings.

In this paper, we fill this gap by presenting a L2O framework for learning fixed points of general nonexpansive mappings with convergence guarantees. Inspired by \cite{martin2025learning} and building on well-known quasi-Fejér monotonicity results for inexact KM iterations, we first show how one can learn convergent NN updates by training summable perturbations to standard KM iterations. Second, under a metric sub-regularity assumption, we prove that our parametrization encompasses only iterations that converge linearly—up to a vanishing bias term—in a neighborhood of a fixed point, and that it includes all iterations that do so at a sufficiently fast rate. Third, we demonstrate how the proposed framework can be used to augment various operator splitting methods for solving structured monotone inclusion problems. Our theoretical results are validated by numerical experiments on a best approximation problem using an L2O-augmented Douglas–Rachford (DR) splitting algorithm \citep{bauschke2013projection}.

\subsection{Notation and Preliminaries}

\emph{Basic notation:} We denote by $\ell^n$ the set of all sequences $\bm{x} = (x^0, x^1, \dots)$ with $x^t \in \R^n$ at all times. For $\bm{x} \in \ell^n$, we let $z\bm{x} = (x^1, x^2, \dots)$ be the sequence shifted one time step forward. We use $\ell^n_1$ to denote the set of all summable sequences, that is, the set of all sequences $\bm{x} \in \ell^n$ such that $\norm{\bm{x}}_1 = \sum_{t=0}^\infty ~ |x^t| < \infty$, where $| \cdot |$ is an arbitrary vector norm. 

\emph{Convex analysis and operator theory:} We mostly adopt standard
operator theoretic notation and definitions from \cite{bauschke2017convex}. The convex conjugate of $f : \R^n \to (-\infty, +\infty]$ is the function $f^* : \R^n \to (-\infty, +\infty]$ defined by $f^*(y) = \sup_{x \in \R^n} ~ \langle y, x\rangle - f(x)$. For a closed convex set $\mathcal{Y} \subseteq \R^n$, we denote the projection of $x \in \R^n$ onto $\mathcal{Y}$ by $\operatorname{proj}_{\mathcal{Y}}(x) = \argmin_{y \in \mathcal{Y}}~|y-x|$ and the distance of $x$ from $\mathcal{Y}$ by $\operatorname{dist}(x, \mathcal{Y}) = \min_{y \in \mathcal{Y}}~|y-x|$. The normal cone of $\mathcal{Y}$ is the operator $\operatorname{N}_{\mathcal{Y}} : \R^n \rightrightarrows \R^n$ defined by $\operatorname{N}_{\mathcal{Y}}(x) = \{v \in \R^n : \sup_{y \in \mathcal{Y}} ~ \langle v, y-x \rangle \leq 0\}$ if $x \in \mathcal{Y}$, and $\operatorname{N}_{\mathcal{Y}}(x) = \emptyset$ otherwise. We denote the graph of an operator $A: \R^n \rightrightarrows \R^n$ by $\operatorname{gra}_A = \{(x, y) \in \R^n \times \R^n : y \in A(x)\}$. We denote the set of zeros of $A$ by $\operatorname{zer}_A = \{x \in \R^n : 0 \in A(x)\}$. We say that $A$ is monotone if $\langle x-y, u-v\rangle \geq 0$ for any $(x, u) \in \operatorname{gra}_A$ and $(y, v) \in \operatorname{gra}_A$; additionally, $A$ is maximal monotone if there is no other monotone operator $B: \R^n \rightrightarrows \R^n$ such that $\operatorname{gra}_A \subset \operatorname{gra}_B$. For any $\beta > 0$, a single-valued operator $A : \R^n \to \R^n$ is $\beta$-cocoercive if $\langle x-y, T(x)-T(y)\rangle \geq \beta |T(x) - T(y)|^2$ for any $x, y \in \R^n$. For any $\gamma > 0$, the resolvent of $A$ is $J_{\gamma A} = (\Id + \gamma A)^{-1}$ and the reflected resolvent of $A$ is $R_{\gamma A} = 2 J_{\gamma A} - \Id$.

\section{Problem Formulation}
We consider parametric fixed-point problems of the form
\begin{equation}
\label{eq:parametric-fixed-point-problem}
    \operatorname{find} ~ x ~ \operatorname{such~that} ~ x = T_\theta(x)\,,
\end{equation}
where $x \in \R^n$ is the decision variable and $\theta \in \Theta \subseteq \R^p$ is a vector of problem parameters defining each instance of (1) via the fixed-point operator $T_\theta$. Throughout the paper, we impose the following standing assumption.

\begin{assum}
\label{assum:nonempty-nonexpansive}
    For every $\theta \in \Theta$, the set of fixed-points $\operatorname{fix}_{T_{\theta}} = \{x \in \R^n : x = T_{\theta}(x)\}$ of $T_{\theta}$ is nonempty and the operator $T_\theta$ is nonexpansive, that is,
    \begin{equation}
    \label{eq:nonexpansive-fixed-point-operator}
        |T_\theta(x) - T_\theta(y)| \leq |x-y|\,, ~ \forall x,y \in \R^n\,.
    \end{equation}
\end{assum}

The condition $\operatorname{fix}_{T_\theta} \neq \emptyset$ guarantees that \eqref{eq:parametric-fixed-point-problem} admits at least one solution for all 
$\theta \in \Theta$. The nonexpansiveness condition \eqref{eq:nonexpansive-fixed-point-operator} imposes 
a uniform $1$-Lipschitz bound on $T_\theta$ and includes, as special cases, contractive and averaged operators, see, e.g., \cite{bauschke2017convex}.

It is well-known that mere nonexpansiveness of $T_{\theta}$ does not guarantee that the Banach–Picard iteration $x^{t+1} = T_{\theta}(x^t)$, initialized with an arbitrary $x^0 \in \R^n$, converges to a solution of \eqref{eq:parametric-fixed-point-problem} as $t \to \infty$.\footnote{This is the case, e.g., when $T_{\theta} = -\operatorname{Id}$, for all $\theta \in \Theta$, and $x_0 \neq 0$.} To address \eqref{eq:parametric-fixed-point-problem}, we therefore consider general time-varying iterations of the form
\begin{equation}
\label{eq:dynamic-iteration}
    x^{t+1} = S_\theta^t(T_{\theta}, x^{t:0})\,, ~ x^0 \in \R^n\,, ~ t \in \N\,,
\end{equation}
where $S_\theta^t$ denotes an update rule that may depend on the fixed-point operator $T_{\theta}$ and on  the past iterates $x^{t:0}$. In particular, we note that considering dynamic iterations as per \eqref{eq:dynamic-iteration} is crucial to describe not only the standard KM iteration \citep{krasnosel1955two, mann1953mean}
\begin{equation}
\label{eq:KM-iteration}
    x^{t+1} = x^t + \lambda (T_\theta(x^t) - x^t)\,, ~ x^0 \in \R^n\,, ~ \forall t \in \N\,,
\end{equation}
but also the Halpern iteration $x^{t+1} = x^0 + \lambda (T_\theta(x^t) - x^0)$ with anchor point $x^0$ \citep{halpern1967}, and several recently proposed accelerated schemes \citep{walker2011anderson, bot2023fast, maulen2024inertial}.

We impose the following two desiderata. First, we require the iteration \eqref{eq:dynamic-iteration} to be asymptotically regular and convergent to $\operatorname{fix}_{T_\theta}$, as per the following definition.
\begin{defn}
\label{def:asymptotic-reg-convergence}
    An iteration of the form \eqref{eq:dynamic-iteration} is asymptotically regular, over the parametrization $\Theta$, if
    \begin{equation}
    \label{eq:asymptotic-regularity}
        \lim_{t \to \infty} ~ |x^t - S^t_\theta(T_\theta, x^{t:0})| = 0\,, ~ \forall x^0 \in \R^n\,, ~ \forall \theta \in \Theta\,.
    \end{equation}
    Additionally, such an iteration is convergent for the parametric fixed-point problem \eqref{eq:parametric-fixed-point-problem} if
    \begin{equation}
    \label{eq:asymptotic-convergence}
        \lim_{t \to \infty} \operatorname{dist}(x^t, \operatorname{fix}_{T_\theta}) = 0\,, ~ \forall x^0 \in \R^n\,, ~ \forall \theta \in \Theta\,.
    \end{equation}
\end{defn}
Second, given a probability distribution $\mathbb{P}_\theta$ over problem instances, we seek update rules $\bm{x} = \bm{S}_{\theta}(T_{\theta}, \bm{x})$, with $\bm{S}_{\theta}(T_{\theta}, \bm{x}) = (S^0_{\theta}(T_\theta, x^0), S^1_{\theta}(T_\theta, x^{1:0}), \dots)$ for compactness, that achieve favorable  average-case performance.

Combining both desiderata, we formalize the problem of designing an optimal solution method for \eqref{eq:parametric-fixed-point-problem} as follows:
\begin{subequations}
\label{eq:optimal-iteration-design-problem}
    \begin{align}
        \label{eq:meta_loss}
         \min_{\bm{S}_{\theta}} &~~ \mathbb{E}_{\mathbb{P}_{\theta}} [\mathtt{MetaLoss}(\bm{S}_{\theta}, T_{\theta}, \bm{x})]\\
        \st &~~ z \bm{x} = \bm{S}_\theta(T_\theta, \bm{x}), \, x^0 \in \R^n\,, \eqref{eq:asymptotic-regularity}\,, \eqref{eq:asymptotic-convergence}\,,    \end{align}
\end{subequations}
where $\mathtt{MetaLoss}(\bm{S}_{\theta}, T_{\theta}, \bm{x})$ represents an algorithmic cost function of choice. For instance, one can let
\begin{equation*}
    \mathtt{MetaLoss}(\bm{S}_{\theta}, T_{\theta}, \bm{x}) = \norm{\bm{x} - \bm{S}_\theta(T_\theta, \bm{x})}_1\,,
\end{equation*}
to measure how fast the residuals of the iteration \eqref{eq:dynamic-iteration} converge to zero for a fixed $\theta \sim \mathbb{P}_{\theta}$.

\section{Main Results}
In this section, we characterize a class of inexact KM iterations in the form \eqref{eq:dynamic-iteration} that guarantee convergence to a solution of \eqref{eq:parametric-fixed-point-problem}, as per Definition~\ref{def:asymptotic-reg-convergence}. Under an additional metric sub-regularity assumption, we then prove that our parametrization encompasses only iterations that converge linearly—up to a vanishing bias term—in a neighborhood of a fixed point, and that it includes all iterations that achieve such convergence at a sufficiently fast rate.
Finally, we show how our results enable learning perturbations to several operator splitting methods to improve their average-case performance while preserving their convergence guarantees. 

\subsection{A characterization of convergent iterations}

Differently from the Banach–Picard iteration $x^t = T_{\theta}(x^t)$, the KM iteration \eqref{eq:KM-iteration} performs convex combinations of the current iterate $x^t$ and its image under $T_\theta$, and generates a sequence $\bm{x}$ that is convergent as per Def.~\ref{def:asymptotic-reg-convergence}. As such, this iteration represents a feasible—yet possibly suboptimal—solution to \eqref{eq:optimal-iteration-design-problem}. Inspired by \cite{martin2024learning} and 
\cite{martin2025learning}, the following proposition shows that different feasible solutions $\bm{S}_\theta(T_{\theta}, \bm{x})$ for \eqref{eq:optimal-iteration-design-problem} can be obtained by perturbing the KM iteration with a summable augmentation term $\bm{v} \in \ell_1^n$.
\begin{prop}
\label{prop:sufficiency}
    For any $\lambda \in (0,1)$ and any $\bm{v} \in \ell^n_1$, the iteration $x^{t+1} = S_\theta^t(T_{\theta}, x^{t:0})$ defined by 
    \begin{equation}
    \label{eq:KM-iteration-plus-v}
        x^{t+1} = x^t + \lambda (T_\theta(x^t) - x^t) + v^t\,, ~ x^0 \in \R^n\,, ~ \forall t \in \N\,,
    \end{equation}
    satisfies \eqref{eq:asymptotic-regularity} and \eqref{eq:asymptotic-convergence}.
\end{prop}
\begin{pf}
    By construction, \eqref{eq:KM-iteration-plus-v} corresponds to an inexact KM iteration. Since $\bm{v} \in \ell^n_1$ by assumption, the claim then follows from quasi-Fejér monotonicity properties of such iterations, see, for instance, Proposition~5.34 in \cite{bauschke2017convex}.
    $\hfill \qed$
\end{pf}

Proposition~\ref{prop:sufficiency} reinterprets the convergence results of inexact KM iterations from a L2O perspective, and suggests treating $v^t$ as an augmentation term that is deliberately introduced to improve the algorithmic performance of the recursion \eqref{eq:KM-iteration-plus-v} rather than as an uncontrollable error term. In Section~\ref{sec:experiments}, we will show how one can leverage NN models to parametrize $\bm{v}$ and learn augmentations that accelerate convergence for specific realizations of $\theta \sim \mathbb{P}_\theta$. 

Next, we investigate the expressiveness of searching over $\bm{v} \in \ell_1^n$ as per \eqref{eq:KM-iteration-plus-v}.

\begin{exmp}
\label{ex:counterexample-completeness}
    Let $T_\theta =0$ be the zero operator. Note that this choice satisfies Assumption~\ref{assum:nonempty-nonexpansive} as, for every $\theta \in \Theta$, the operator $T_\theta$ is nonexpansive and $\operatorname{fix}_{T_\theta} = \{0\} \neq \emptyset$. For any $z^0 \in \R^n$, consider the iteration $z^{t+1} = \hat{S}_\theta^t(T_{\theta}, z^{t:0})$ defined by $z^{t} = \frac{1}{t} z^0$. Note that $\hat{\bm{S}}$ is asymptotically regular and convergent as per Definition~\ref{def:asymptotic-reg-convergence}, and is thus a feasible solution for \eqref{eq:optimal-iteration-design-problem}. However, the sequence $\bm{z}$ generated by $\hat{\bm{S}}$ cannot be decomposed as per \eqref{eq:KM-iteration-plus-v} with $x^0 = z^0$ and $\bm{v} \in \ell_1^n$. In fact, assuming that $x^k = z^k$ for all $0 \leq k \leq t$ as inductive hypothesis, one would need to select
    \begin{align}
        \label{eq:v_matching_sequence}
        v^t &= \hat{S}_\theta^t(T_{\theta}, z^{t:0}) - (x^t + \lambda(T_\theta(x^t) - x^t))\\
        \nonumber
        &= \frac{z^0}{t+1} - (1 - \lambda) x^t = z^0\frac{\lambda t + \lambda - 1}{t(t+1)}\,, 
    \end{align}
    to ensure that $x^{t+1} = z^{t+1}$. Hence, for sufficiently large $t \in \N$, $v^t = \frac{\lambda z^0}{t} + \mathcal{O}\left(\frac{1}{t^2}\right)$ and thus $\bm{v} \not \in \ell_1^n$ unless $z^0 = 0$.
\end{exmp}

Differently from the case of linearly convergent operators studied in \cite{martin2025learning}, the counterexample above shows that not all iterations $\bm{S}_\theta$ complying with \eqref{eq:asymptotic-regularity} and \eqref{eq:asymptotic-convergence} correspond to a specific choice of $\bm{v} \in \ell_1^n$ in \eqref{eq:KM-iteration-plus-v}. This is because the condition $\bm{v} \in \ell_1^n$ imposes a superlinear decay of $v^t$, yet the iterations $\bm{S}_\theta$ in the feasible set of \eqref{eq:optimal-iteration-design-problem} can exhibit arbitrarily slow convergence to $\operatorname{fix}_{T_{\theta}}$. 

We proceed to show that the proposed parametrization is nevertheless highly expressive, as, when the following standard regularity condition holds \citep{dontchev2009implicit}, it encompasses all—and, up to a vanishing bias term, only—the iterations $\bm{S}_\theta$ that achieve sufficiently fast local linear convergence.

\begin{assum}
\label{ass:metric_sub_regularity}
    The operator $R_{\theta} = \Id - T_\theta$ is \emph{metric sub-regular} at some $x_\theta^\star \in \operatorname{fix}_{T_\theta}$ for $0$, that is, there exists $\kappa_\theta \in \mathbb{R}$, $\kappa_\theta \geq 0$, and a neighborhood $\mathcal{X}_\theta$ of $x_\theta^\star$ such that
    \begin{equation}
    \label{eq:metric_sub_regularity}
        \operatorname{dist}(x, \operatorname{fix}{T_\theta}) \leq \kappa_\theta |R_\theta x| = \kappa_\theta |x - T_\theta (x)| \,, ~ \forall x \in \mathcal{X}_\theta\,.
    \end{equation}
\end{assum}

Metric sub-regularity allows one to estimate how far an iterate $x^t$ is from the set of fixed-points of $T_\theta$ in terms of the fixed-point residual $|x^t - T_\theta(x^t)|$. Many operators $T_{\theta}$ satisfy \eqref{eq:metric_sub_regularity}; for instance, if $T_\theta = \operatorname{proj}_{\mathcal{C}_{\theta}}$ is the projection operator on a nonempty, closed, and convex set $\mathcal{C}_{\theta} \subseteq \R^n$, then $\operatorname{fix}_{T_\theta} = \mathcal{C}_{\theta}$ and \eqref{eq:metric_sub_regularity} holds with $\kappa_\theta = 1$ and $\mathcal{X}_{\theta} = \R^n$ since $\operatorname{dist}(x, \mathcal{C}_{\theta}) = |x - \operatorname{proj}_{\mathcal{C}_\theta}(x)|$.

We are ready to present our local completeness result.

\begin{thm}
\label{th:local-necessity}
    Let Assumption~\ref{ass:metric_sub_regularity} hold and fix any nonnegative constants $r_0$ and $r_v$ such that $\mathbb{B}_{r_0 + r_v}(x_\theta^\star) = \{x \in \R^n : |x - x^\star_\theta| \leq r_0 + r_v\}\subseteq \mathcal{X}_\theta$. The following statements hold:
    \begin{enumerate}[leftmargin=*]
        \item[\emph{i})] For any $x^0 \in \mathbb{B}_{r_0}(x_\theta^\star)$ and $\bm{v} \in \ell_1^n$ with $\norm{\bm{v}}_1 \leq r_v$, the iteration \eqref{eq:KM-iteration-plus-v} satisfies $x^t \in \mathbb{B}_{r_0 + r_v}(x_\theta^\star)$ at all times and there exists $\nu_\theta \in \R$ such that
        \begin{equation}
        \label{eq:quasi-linear-convergence-inexact-km}
            \operatorname{dist}^2(x^{t+1}, \operatorname{fix}_{T_\theta}) \leq \zeta_\theta \operatorname{dist}^2(x^{t}, \operatorname{fix}_{T_\theta}) + \nu_\theta |v^t|\,,%
        \end{equation}
        where $\zeta _\theta= \frac{\kappa_\theta^2}{\kappa_\theta^2 + \lambda (1 - \lambda)} \in (0, 1)$, for every $t \in \N$.
        \item[\emph{ii})] Let $z^{t+1} = \hat{S}_\theta^t(T_{\theta}, z^{t:0})$ be any iteration such that, for any $z^0 \in \mathbb{B}_{r_0}(x_\theta^\star)$ and $\theta \in \Theta$,
        \begin{align*}
            \operatorname{dist}(z^{t}, \operatorname{fix}_{T_\theta}) \leq c\rho^t\,, ~ |z^t - \hat{S}_\theta^t(T_{\theta}, z^{t:0})| \leq c \rho^t\,, ~ \forall t \in \N\,,   
        \end{align*}
        for some $c > 0$ and $\rho \in (0,1)$. If $3c < r_v (1-\rho)$, then there exists $\bm{v}$ with $\norm{\bm{v}}_1 < r_v$ such that the sequence $\bm{x}$ defined by \eqref{eq:KM-iteration-plus-v} with $x^0 = z^0$ is equivalent to $\bm{z}$.
    \end{enumerate}     
\end{thm}

\begin{pf}
    By interpreting the recursion \eqref{eq:KM-iteration-plus-v} as an inexact KM iteration and observing that \eqref{eq:KM-iteration-plus-v} matches \cite[Equation~4]{liang2016convergence} after the change of variables $v^t = \lambda \epsilon^t$, the bound \eqref{eq:quasi-linear-convergence-inexact-km} in $i)$ follows from the (not necessarily tight) convergence rates established in Theorem~3 of \cite{liang2016convergence}. We prove \emph{ii}). As $x^0 = z^0$ by assumption, we can construct a sequence $\bm{v}$ that ensures $x^t = z^t$ at all times in the same way as in Example~\ref{ex:counterexample-completeness}. It remains to show that the sequence $\bm{v}$ in \eqref{eq:v_matching_sequence} satisfies $\bm{v} \in \ell_1^n$ and $\norm{\bm{v}}_1 \leq r_v$. By introducing the $\lambda$-averaged operator $T_\theta^\lambda = (1 - \lambda) \Id + \lambda T_\theta$ and letting $\bar{z}^t = \argmin_{z \in \operatorname{fix}_{T_{\theta}}} ~ |z^t - z|$,\footnote{Note that the set $\operatorname{fix}_{T_\theta}$ is closed and convex since $T_\theta$ is nonexpansive \citep[Corollary~4.24]{bauschke2017convex}. Hence, the projection onto $\operatorname{fix}_{T_\theta}$ always exists and is unique.} we rewrite \eqref{eq:v_matching_sequence} as
    \begin{equation*}
        v^t = \hat{S}_\theta^t(T_{\theta}, z^{t:0}) - z^t - T_{\theta}^\lambda(z^t) + \bar{z}^t + z^t - \bar{z}^t\,.
    \end{equation*}
    Using the triangle inequality, we then have that
    \begin{align*}
        |v^t| &\leq |\hat{S}_\theta^t(T_{\theta}, z^{t:0}) - z^t| + |T_{\theta}^\lambda(z^t) - \bar{z}^t| + |z^t - \bar{z}^t|\\
        &\overset{(a)}{=} |\hat{S}_\theta^t(T_{\theta}, z^{t:0}) - z^t| + |T_{\theta}^\lambda(z^t) - T_{\theta}^\lambda(\bar{z}^t)| + |z^t - \bar{z}^t|\\
        &\overset{(b)}{\leq} |\hat{S}_\theta^t(T_{\theta}, z^{t:0}) - z^t| + 2 |z^t - \bar{z}^t|\\
        &= |\hat{S}_\theta^t(T_{\theta}, z^{t:0}) - z^t| + 2\operatorname{dist}(z^t, \operatorname{fix}_{T_\theta}) \leq 3c \rho^t\,,
    \end{align*}
    where $(a)$ follows from the fact that $T_{\theta}$ and $T^\lambda_{\theta}$ share the same set of fixed-points and $(b)$ from nonexpansiveness of $T_\theta^\lambda$. We therefore have that $\norm{\bm{v}}_1 = \sum_{t = 0}^{\infty} ~ |v^t| \leq \frac{3c}{1 - \rho}$, which implies the bound $\norm{\bm{v}}_1 < r_v$ since $3c < r_v(1-\rho)$ by assumption. This concludes the proof.
    $\hfill \qed$
\end{pf}

It is well-known that Assumption~\ref{ass:metric_sub_regularity} suffices to prove local linear convergence of the standard KM iteration \citep{bauschke2015linear}. Building on  \cite{liang2016convergence}, Theorem~\ref{th:local-necessity} bounds the effect of injecting a summable perturbation in \eqref{eq:KM-iteration-plus-v} on this local convergence guarantee by means of the vanishing bias term $\nu_{\theta} |v^t|$ in \eqref{eq:quasi-linear-convergence-inexact-km}. In addition, Theorem~\ref{th:local-necessity} shows that all iterations $\bm{S}_\theta$ that locally converge linearly with a sufficiently fast rate can be parametrized as per \eqref{eq:KM-iteration-plus-v} by accurately selecting $\bm{v} \in \ell_1^n$.

\subsection{Application to operator splitting methods}
We now show how the results of Proposition~\ref{prop:sufficiency} and Theorem~\ref{th:local-necessity} provide a principled way to inject trainable perturbations into the Davis–Yin splitting \citep{davis2017three}, a widely-used decomposition method that recovers many other splittings as special cases.

We consider monotone inclusion problems of the form
\begin{equation}
\label{eq:monotone-inclusion}
    \operatorname{find} ~ x \in \R^n ~ \operatorname{such~that} ~ 0 \in A(x) + B(x) + C(x)\,,
\end{equation}
where $A: \R^n \rightrightarrows \R^n$ and $B: \R^n \rightrightarrows \R^n$ are set-valued maximal monotone operators, and $C: \R^n \to \R^n$ is single-valued, maximal monotone, and $\beta$-cocoercive. To address this problem, the Davis–Yin splitting \citep{davis2017three} performs a backward step on $B$, followed by a forward step on $C$ and a backward step on $A$, yielding
\begin{subequations}
    \label{eq:davis-yin-splitting}
    \begin{align}
        x^t &= J_{\gamma B}(z^t)\\
        y^t &= J_{\gamma A}(2x^t - z^t - \gamma C(x^t))\\
        z^{t+1} &= z^t + y^t - x^t\,,
    \end{align}
\end{subequations}
where $\gamma > 0$ is a design parameter. In particular, we remark that \eqref{eq:davis-yin-splitting} reduces, for instance, to the backward-forward splitting when $A = 0$, to the forward-backward splitting when $B = 0$, and to the DR splitting when $C = 0$, see \cite{bauschke2017convex}.

\begin{thm}
\label{th:davis-yin-plus-v-convergence}
    For any $\gamma \in (0, 2\beta)$ and any $\bm{v}_x, \bm{v}_y, \bm{v}_z \in \ell_1^n$, the iteration defined by
    \begin{subequations}
    \label{eq:davis-yin-splitting-plus-v}
        \begin{align}
            \label{eq:davis-yin-splitting-plus-v-update-x}
            x^t &= J_{\gamma B}(z^t) + v_x^t\\
            \label{eq:davis-yin-splitting-plus-v-update-y}
            y^t &= J_{\gamma A}(2x^t - z^t - \gamma C (x^t)) + v_y^t\\
            \label{eq:davis-yin-splitting-plus-v-update-z}
            z^{t+1} &= z^t + y^t - x^t + v_z^t\,,
        \end{align}
    \end{subequations}
    converges to a solution of \eqref{eq:monotone-inclusion} if $\operatorname{zer}_{A + B + C} \neq \emptyset$.
\end{thm}
\begin{pf}
    By plugging \eqref{eq:davis-yin-splitting-plus-v-update-x} and \eqref{eq:davis-yin-splitting-plus-v-update-y} into \eqref{eq:davis-yin-splitting-plus-v-update-z} and after some algebraic manipulations, we first compactly rewrite the fixed-points iteration \eqref{eq:davis-yin-splitting-plus-v} as 
    \begin{equation}
    \label{eq:davis-yin-splitting-plus-v-compact}
        z^{t+1} = T_{DY}(z^t) + v^t\,,
    \end{equation}
    where $T_{DY} = \Id - J_{\gamma B} + J_{\gamma A} \circ (2R_{\gamma B} - \gamma C \circ J_{\gamma B})$ is the fixed-point operator associated with \eqref{eq:davis-yin-splitting} and
    \begin{align}
        \nonumber   
        v^t &= v_z^t + v_y^t - v_x^t - J_{\gamma A}\left(R_{\gamma B}(z^t) - \gamma C(J_{\gamma B}(z^t))\right)\\
        \label{eq:dys-v-compact}
        &\quad + J_{\gamma A}\left(R_{\gamma B}(z^t) + v_x^t - \gamma C(J_{\gamma B}(z^t + v_x^t))\right)\,. 
    \end{align}
    Recall that the operator $T_{DY}$ in \eqref{eq:davis-yin-splitting-plus-v-compact} is $\alpha$-averaged with $\alpha = \frac{2\beta}{4\beta - \gamma}$ for any $\gamma \in (0, 2\beta)$, see Theorem 28 in \S 10 of \cite{ryu2022large}. Hence, there exists a nonexpansive operator $R_{DY}$ such that $T_{DY} = (1-\alpha) \Id + \alpha R_{DY}$ and, by Proposition~\ref{prop:sufficiency}, we have that \eqref{eq:davis-yin-splitting-plus-v-compact} converges to $\operatorname{fix}_{T_{DY}} = \operatorname{fix}_{R_{DY}}$—and therefore \eqref{eq:davis-yin-splitting-plus-v-update-x} converges to a solution of \eqref{eq:monotone-inclusion}—if the sequence $\bm{v}$ defined by \eqref{eq:dys-v-compact} is summable.
    
    We proceed to show that $\bm{v} \in \ell_1^n$. By the triangle inequality and the fact that the resolvent of a maximal monotone operator is (firmly) nonexpansive \citep[Corollary~23.9]{bauschke2017convex}, we can upper bound $|v^t|$ by
    \begin{equation*}
        |v_z^t| + |v_y^t| + 2|v_x^t| + \gamma |C(J_{\gamma B}(z^t + v_x^t)) - C(J_{\gamma B}(z^t))|\,.
    \end{equation*}
    Moreover, since $\beta$-cocoercivity implies $\frac{1}{\beta}$-Lipschitzness \citep[Remark~4.15]{bauschke2017convex}, we obtain
    \begin{align*}
        |v^t| &\leq |v_z^t| + |v_y^t| + 2|v_x^t| + \frac{\gamma}{\beta} |J_{\gamma B}(z^t + v_x^t) - J_{\gamma B}(z^t)|\\
        &\leq |v_z^t| + |v_y^t| + (2 + \gamma \beta^{-1})|v_x^t|\,,
    \end{align*}
    from which the proof follows since $\bm{v}_x, \bm{v}_y, \bm{v}_z \in \ell_1^n$.
    $\hfill \qed$
\end{pf}

Theorem~\ref{th:davis-yin-plus-v-convergence} demonstrates that introducing summable perturbations in the updates of \eqref{eq:davis-yin-splitting-plus-v} preserves the convergence guarantees of \eqref{eq:davis-yin-splitting}. In particular, differently from related results for the forward–backward splitting in \cite{sadeghi2024incorporating}, our convergence guarantees do not impose any instantaneous norm bounds on $\bm{v}_x, \bm{v}_y$ and $\bm{v}_z$.

\begin{rem}
    Our results naturally extend to augmenting any algorithm that can be derived as a special case of the Davis–Yin splitting \eqref{eq:davis-yin-splitting}. For instance, the well-known alternating direction method of multipliers (ADMM) algorithm for solving optimization problems of the form
    \begin{equation*}
        \min_{x, z} ~ f(x) + g(z) ~ \st ~ Fx + Gz = h\,,
    \end{equation*}
    where $f$ and $g$ are closed convex proper functions, can be obtained by applying the DR splitting to the dual monotone inclusion 
    \begin{equation*}
        0 \in \partial f^*(-F^\top u) + \partial g^*(-G^\top u) + h\,,
    \end{equation*}
    which is the optimality condition of the dual problem
    \begin{equation*}
        \min_{u} ~ f^*(-F^\top u) + g^*(-G^\top u) + \langle h, u\rangle\,;
    \end{equation*}
    see, for instance, \cite{eckstein1992douglas} and \cite{ryu2022large} for a more detailed derivation.
\end{rem}

In the next section, we will show how one can train NN updates to accelerate convergence of \eqref{eq:davis-yin-splitting-plus-v} to a solution of \eqref{eq:monotone-inclusion} by learning parametrized perturbations $\bm{v} \in \ell_1^n$. 

\section{Experiments}
\label{sec:experiments}

\begin{figure*}[htb]
\centering
\begin{subfigure}{\columnwidth}
  \centering
  \includegraphics[width=\linewidth]{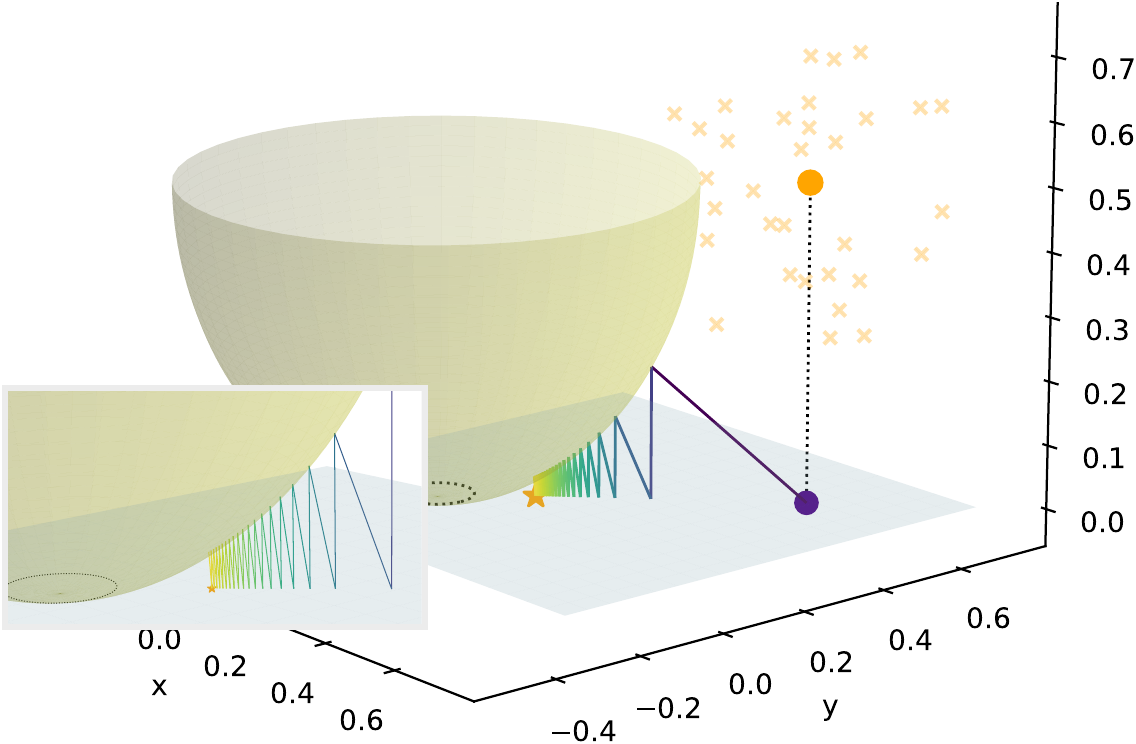}
  \caption{Best approximation Douglas–Rachford (baDR).}
  \label{fig:shadow-iterates-baDR}
\end{subfigure}%
\hfill
\begin{subfigure}{\columnwidth}
  \centering
  \includegraphics[width=\linewidth]{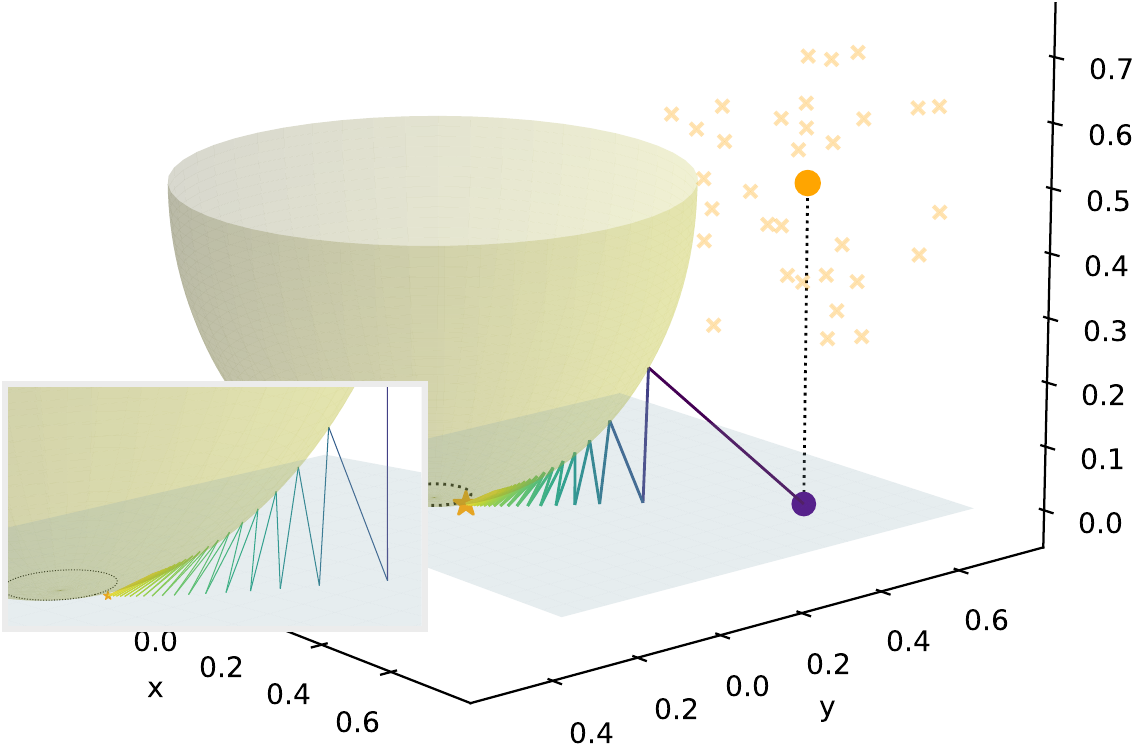}
  \caption{Ours (augmented baDR).}
  \label{fig:shadow-iterates-baDR-l2o}
\end{subfigure}
\caption{Comparison between the evolution of the first $20$ shadow iterates $x^t$ and $y^t$ generated by \eqref{eq:baDR} and \eqref{eq:baDR-l2o}. The orange point indicates the query point to be projected, and the shaded orange crosses denote the training samples.}
\label{fig:comparison-shadow-iterates}
\end{figure*}

We now illustrate how the results of Theorem~\ref{th:davis-yin-plus-v-convergence} allow learning augmentations that improve the average-case performance of an operator splitting method without sacrificing its convergence guarantees.\footnote{The source code used to reproduce our numerical experiments is available at \href{https://github.com/andrea-martin/l2o-nonexp-fpi}{github.com/andrea-martin/l2o-nonexp-fpi}.} We consider the classical problem of computing the projection of a point $p \in \R^n$ onto the intersection of two nonempty closed convex sets $\mathcal{C}_1 \subseteq \R^n$ and $\mathcal{C}_2 \subseteq \R^n$. This leads us to the best-approximation problem
\begin{equation}
\label{eq:projecting-intersection-problem}
    \operatorname{find} ~ x \in \R^n ~ \operatorname{such~that} ~ x =\argmin_{y \in \mathcal{C}_{1} \cap \mathcal{C}_2} ~ \frac{1}{2}|y - p|^2\,.
\end{equation}
To address this problem, a standard approach is to reformulate \eqref{eq:projecting-intersection-problem} as the monotone inclusion problem
\begin{equation}
    \label{eq:monotone-inclusion-projecting-intersection}
    \operatorname{find} ~ x \in \R^n ~ \operatorname{such~that} ~ 0 \in x - p + \operatorname{N}_{\mathcal{C}_1}(x) + \operatorname{N}_{\mathcal{C}_2}(x)\,,
\end{equation}
and then apply the DR splitting to the maximal monotone operators $A(x) = x - p + \operatorname{N}_{\mathcal{C}_1}(x)$ and $B(x) = \operatorname{N}_{\mathcal{C}_2}(x)$.\footnote{Recall that the subdifferential of any closed convex proper function is maximal monotone, and so is the sum of two maximal monotone operators $A_1$ and $A_2$ satisfying $\operatorname{dom}_{A_1} \cap \operatorname{int~dom}_{A_2} \neq \emptyset$, see \cite{ryu2022large}.} This leads to the DR iterations 
\begin{subequations}
\label{eq:baDR}
    \begin{align}
        \label{eq:baDR-update-x}
        x^t &= \operatorname{proj}_{\mathcal{C}_2}(z^t)\\
        \label{eq:baDR-update-y}
        y^t &= \operatorname{proj}_{\mathcal{C}_1}\left(x^t - \frac{1}{2} z^t + \frac{1}{2} p\right)\\
        \label{eq:baDR-update-z}
        z^{t+1} &= z^t + y^t - x^t\,,
    \end{align}
\end{subequations}
see \citep{bauschke2013projection} for a complete derivation. 

In our experiments, we consider an instance of \eqref{eq:projecting-intersection-problem} where $\mathcal{C}_1$ is a ball and $\mathcal{C}_2$ is a linear subspace that is nearly parallel to the tangent space to the boundary sphere $\partial \mathcal{C}_1$ at any point in $\partial \mathcal{C}_1 \cap \mathcal{C}_2$. We assume that the points $p \in \R^3$ are randomly generated from a uniform distribution $\mathbb{P}$ over the cube $[0.25, 0.75]^3$, and define \eqref{eq:meta_loss} as the expected weighted sum of the residuals $|z^{t+1} - z^t|$ over $t \in \N$ when $p \sim \mathbb{P}$. 
To improve the average-case performance of \eqref{eq:baDR} in projecting points $p \sim \mathbb{P}$, we introduce a trainable perturbation $v_z^t$ in \eqref{eq:baDR-update-z}, resulting in the iterations
\begin{subequations}
\label{eq:baDR-l2o}
    \begin{align}
        \label{eq:baDR-l2o-update-x}
        x^t &= \operatorname{proj}_{\mathcal{C}_2}(z^t)\\
        \label{eq:baDR-l2o-update-y}
        y^t &= \operatorname{proj}_{\mathcal{C}_1}\left(x^t - \frac{1}{2} z^t + \frac{1}{2} p\right)\\
        \label{eq:baDR-l2o-update-z}
        z^{t+1} &= z^t + y^t - x^t + v_z^t\,,
    \end{align}
\end{subequations}
To ensure that $\bm{v}_z \in \ell_1^3$ according to Theorem~\ref{th:davis-yin-plus-v-convergence}, we proceed similarly to \cite{martin2025learning} and parametrize $v_z^t$ as the product of an exponentially decaying scalar sequence $m^t$ and a bounded NN update $\delta^t$ encoding the direction of $v_z^t$. Specifically, we let $v_z^t = m^t(\varphi_m) \operatorname{tanh}(\delta^t(z^{t:0}, \varphi_\delta))$, where $m^t(\varphi_m) = \varphi_m \rho^t$, $\rho \in (0,1)$, and $\delta^t$ is the output vector of a two-layer long short-term memory (LSTM) network. To train the parameters $\varphi = (\varphi_m, \varphi_\delta)$, we draw $N = 32$ training points $p_i \sim \mathbb{P}$, $i \in \{1, \dots, N\}$, and use Adam \citep{kingma2014adam} to minimize the sample average approximation of \eqref{eq:meta_loss} for $100$ epochs. With our design choices, this corresponds to minimizing the empirical weighted sum of residuals 
\begin{equation}
\label{eq:meta-loss-saa}
    \frac{1}{N} \sum_{i = 1}^N \sum_{t = 1}^{T} \alpha^{T-t} |z^{t}(p_i, \varphi) - z^{t-1}(p_i, \varphi)|\,,
\end{equation}
where $T = 200$ is the number of iterations over which we unroll \eqref{eq:baDR-l2o} and $\alpha \in (0,1)$ is a discount factor.\footnote{We refer the interested reader to \cite{sambharya2025data} for a discussion of how PAC–Bayes bounds can be used to quantify the gap between the sample average approximation in \eqref{eq:meta-loss-saa} and the true expectation in \eqref{eq:meta_loss} in the context of L2O.}

\begin{figure}[htb]
\centering
\includegraphics[width=\columnwidth]{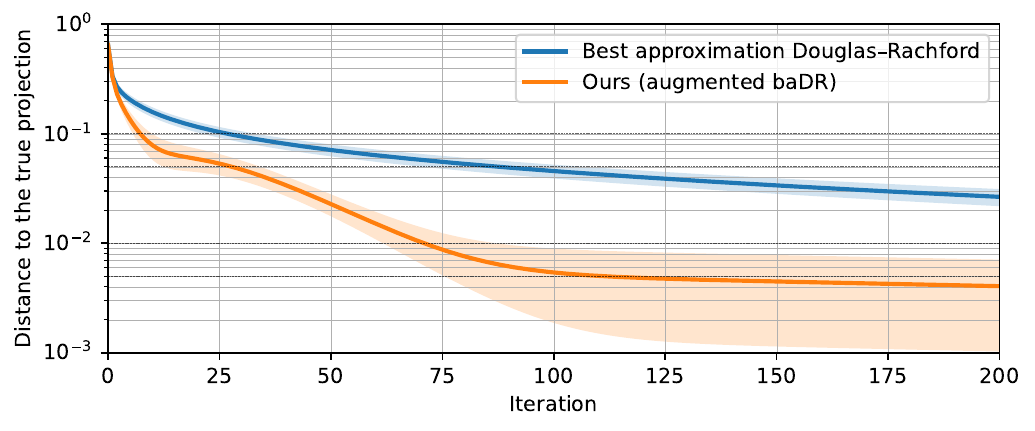}
\caption{Evolution of the distance between the solution of \eqref{eq:projecting-intersection-problem} and the shadow iterates $x^t$ generated by \eqref{eq:baDR} and \eqref{eq:baDR-l2o}. Shaded areas and solid lines denote standard deviations and mean values, respectively.
}
\label{fig:comparison-distance-true-projection}
\end{figure}

In Figure~\ref{fig:comparison-shadow-iterates}, we compare the evolution of the first $20$ shadow iterates $x^t$ and $y^t$ generated by the best approximation DR splitting and its L2O-augmented counterpart through \eqref{eq:baDR} and \eqref{eq:baDR-l2o}, respectively. We observe that, as the plane $\mathcal{C}_2$ is almost parallel to the tangent space to the ball $\mathcal{C}_1$ at any point in the intersection, the approximation DR method \eqref{eq:baDR} progresses very slowly toward the intersection. Conversely, Figure~\ref{fig:comparison-shadow-iterates} shows that, thanks to the augmentations $v_z^t$ in \eqref{eq:baDR-l2o-update-z}, the proposed method learns to slide along the surface of the ball, achieving modest accuracy solutions after a small number of iterations only. This observation is further validated by Figure~\ref{fig:comparison-distance-true-projection}, where we report the average distance between the solution of \eqref{eq:projecting-intersection-problem} and the shadow iterates $x^t$ in \eqref{eq:baDR-update-x} and \eqref{eq:baDR-l2o-update-x} computed over $1024$ independent realizations of $p \sim \mathbb{P}$.

\section{Conclusion}
We have presented a L2O framework for accelerating the convergence of classical KM iterations to the fixed points of general nonexpansive mappings. Specifically, leveraging well-known quasi-Fejér monotonicity properties of inexact KM iterations, we have proposed learning parametrized summable perturbations to standard KM iterations to enhance their average-case performance on a set of training problems through automatic differentiation while guaranteeing convergence by design. We have formally analyzed the expressiveness of the proposed parametrization under a mild regularity assumption, and we have showcased the broad applicability and potential of our framework by augmenting several widely-used operator splitting methods. Important avenues for future research include extensions to stochastic fixed-point iterations and applications to equilibrium-seeking algorithms in multi-player games.

\bibliography{ifacconf}             %

\end{document}